\documentclass[aps,prb,reprint,amsmath,amssymb,superscriptaddress]{revtex4-1}

\usepackage{graphicx}
\usepackage{dcolumn}
\usepackage{bm}

\begin{document}

\title{
  Influence of lattice structure
  on multipole interactions\\
  in $\Gamma_3$ non-Kramers doublet systems
}

\author{Katsunori Kubo}
\affiliation{
  Advanced Science Research Center, Japan Atomic Energy Agency,
  Tokai, Ibaraki 319-1195, Japan}

\author{Takashi Hotta}
\affiliation{
  Department of Physics, Tokyo Metropolitan University,
  Hachioji, Tokyo 192-0397, Japan}

\date{\today}

\begin{abstract}
We study the multipole interactions
between $f^2$ ions
with the $\Gamma_3$ non-Kramers doublet ground state
under a cubic crystalline electric field.
We construct the $\Gamma_3$ doublet state
of the electrons with the total angular momentum $j=5/2$.
By applying the second-order perturbation theory
with respect to the intersite hopping,
we derive the multipole interactions.
We obtain a quadrupole interaction for a simple cubic lattice,
an octupole interaction for a bcc lattice,
and both quadrupole and octupole interactions for an fcc lattice.
We also discuss general tendencies of the multipole interactions
depending on the lattice structure
by comparing the results with those for the $\Gamma_8$ quartet systems
of $f^1$ ions.
\end{abstract}

\maketitle

\section{Introduction}
In the field of the $f$-electron systems,
the phenomena which originate from the multipole degrees of freedom
have been studied intensively
since such degrees of freedom, in addition to the dipole,
are expected to become sources of exotic ordering and physical properties.
The quadrupole moment couples to the lattice and its influence can be detected,
for example, by ultrasonic measurements.
In recent years, even the effects of the octupole moment
have been investigated.

One of the most representative phenomena discovered in multipole physics
is the quadrupole and octupole ordering
in NpO$_2$~\cite{Paixao2002,Tokunaga2005,
  Kubo2005PRBR,Kubo2005PRB,Kubo2005PRBB}
and Ce$_x$La$_{1-x}$B$_6$.~\cite{Akatsu2003,Kubo2003,Kubo2004,
  Morie2004,Mannix2005,Kuwahara2007,Inami2014}
While it is in general difficult to detect the octupole moment,
resonant x-ray scattering,~\cite{Paixao2002,Mannix2005}
NMR,~\cite{Tokunaga2005}
anisotropic magnetization,~\cite{Morie2004}
and neutron scattering~\cite{Kuwahara2007}
experiments have confirmed the octupole order.
In these compounds,
the crystalline electric field (CEF) ground state is the $\Gamma_8$ quartet,
which has sufficient degrees of freedom
to possess quadrupole and octupole moments in addition to the dipole moment.
Then, the $\Gamma_8$ quartet has been regarded
as an ideal system for multipole physics.

However, large degeneracy, such as in a quartet,
is not a necessary condition
to possess higher-order multipole moments.
If the CEF ground state does not have the dipole moment but is not a singlet,
this state inevitably has the higher-order multipole degrees of freedom.
In fact, the $\Gamma_3$ doublet state under a cubic CEF,
which we will explore in this paper,
does not have the dipole
but has the quadrupole moments
$O^0_2$ and $O^2_2$ with the $\Gamma_{3g}$ symmetry
and the octupole moment $T_{xyz}$ with the $\Gamma_{2u}$ symmetry.
The absence of the dipole moment is also an advantage of the $\Gamma_3$ systems
since we can focus only on the higher-order multipoles.
In the $\Gamma_3$ state, the degeneracy is not due to the Kramers theorem,
which is applicable only to an ion with odd number of $f$ electrons.
Thus, we consider an ion with even number of $f$ electrons.

In particular, Pr$^{3+}$ ion has two $f$ electrons
and in some Pr compounds, the CEF ground state is the $\Gamma_3$ doublet.
In recent years, interesting phenomena,
which probably originate from the multipole degrees of freedom,
have been reported for Pr compounds with the $\Gamma_3$ CEF ground state.
In PrPb$_3$, incommensurate quadrupole ordering
has been reported.~\cite{Onimaru2005}
PrIr$_2$Zn$_{20}$ and some of other Pr compounds
with the same crystal structure (Pr 1-2-20 compounds)
with the $\Gamma_3$ CEF ground state
become superconducting at low temperatures,~\cite{Onimaru2010,Sakai2012,
  Matsubayashi2012,Onimaru2012,Tsujimoto2014,Onimaru2016}
which might be mediated by multipole fluctuations.

In this paper,
to elucidate multipole phenomena of the $\Gamma_3$ systems,
we derive the multipole interactions
from a simple model only with $f$-$f$ direct hopping.
While the actual exchange process would be through orbitals
other than the $f$ orbital,
such a process can be represented by effective $f$-$f$ hopping.
An important point is that
the symmetry of the $f$ orbital restricts the form of the hopping
for both cases of the direct and effective hopping
and we will obtain qualitatively the same results
for the multipole interactions.~\cite{Kubo2005PRBB}

The anisotropy in the multipole moments are closely tied
to the real space direction
and the multipole interactions are intrinsically anisotropic.
It is in sharp contrast to the isotropic spin-spin interaction
in a system without spin-orbit coupling.
Thus, the nature of the multipole interactions can depend drastically
on lattice structure.
In the present study,
we pay attention to this point of the multipole interactions.
Then, we derive the multipole interactions
for simple cubic (sc), bcc, and fcc lattices.
We also compare the results of the present model
for the $f^2$-$\Gamma_3$ systems
with those of the $\Gamma_8$ model
for the $f^1$ systems~\cite{Kubo2005PRBR,Kubo2005PRB}
to find common features among these two classes.

\section{Ground and intermediate states}\label{model}
\begin{figure}
  \includegraphics[width=0.99\linewidth]
  {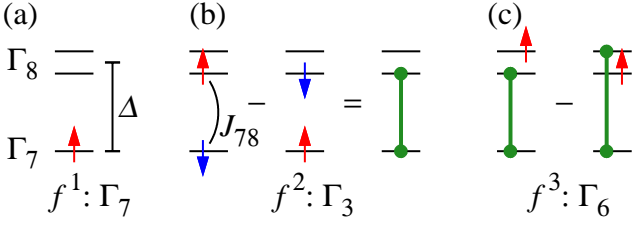}
  \caption{\label{level_scheme}
    (Color online)
    Electron configurations for
    (a) $\Gamma_7$ state of $f^1$,
    (b) $\Gamma_3$ state of $f^2$, and
    (c) $\Gamma_6$ state of $f^3$.
    The bold lines denote spin singlets
    composed of the $\Gamma_7$ and $\Gamma_8$ orbitals.
  }
\end{figure}

To construct electronic states,
we first include the effect of the spin-orbit coupling
in the one-electron states
and consider only the $f$-electron states
with the total angular momentum $j=5/2$.
These states split into the states with $\Gamma_7$ and $\Gamma_8$ symmetry
under a cubic CEF [see Fig.~\ref{level_scheme}(a)].
The $\Gamma_7$ states at site $\bm{r}$ are given by
\begin{subequations}
\begin{align}
  c^{\dagger}_{\bm{r} 7 \uparrow} |0 \rangle
  &\equiv 
  \frac{1}{\sqrt{6}}
  \left(a^{\dagger}_{\bm{r} 5/2}
  -\sqrt{5} a^{\dagger}_{\bm{r} -3/2} \right)|0 \rangle,\\
  c^{\dagger}_{\bm{r} 7 \downarrow} |0 \rangle
  &\equiv 
  \frac{1}{\sqrt{6}}
  \left(a^{\dagger}_{\bm{r} -5/2}
  -\sqrt{5} a^{\dagger}_{\bm{r} 3/2} \right)|0 \rangle,
\end{align}
\end{subequations}
where
$a^{\dagger}_{\bm{r} j_z}$ is the creation operator
of the electron with the $z$-component $j_z$ of the total momentum at $\bm{r}$
and $|0\rangle$ denotes the vacuum state.
The $\Gamma_8$ states are given by
\begin{subequations}
\begin{align}
  c^{\dagger}_{\bm{r} \alpha \uparrow} |0 \rangle
  &\equiv 
  \frac{1}{\sqrt{6}}
  \left( \sqrt{5}a^{\dagger}_{\bm{r} 5/2}
  +a^{\dagger}_{\bm{r} -3/2} \right)|0 \rangle,\\
  c^{\dagger}_{\bm{r} \alpha \downarrow} |0 \rangle
  &\equiv 
  \frac{1}{\sqrt{6}}
  \left( \sqrt{5}a^{\dagger}_{\bm{r} -5/2}
  +a^{\dagger}_{\bm{r} 3/2} \right)|0 \rangle,\\
  c^{\dagger}_{\bm{r} \beta \uparrow} |0 \rangle
  &\equiv
  a^{\dagger}_{\bm{r} 1/2}|0 \rangle,\\
  c^{\dagger}_{\bm{r} \beta \downarrow} |0 \rangle
  &\equiv
  a^{\dagger}_{\bm{r} -1/2}|0 \rangle.
\end{align}
\end{subequations}
In the above equations,
$\sigma=\uparrow$ or $\downarrow$ denotes the Kramers degeneracy
of the one-electron states,
while it is not a real spin because of the spin-orbit coupling.
In the following, however, we call it spin for simplicity.

In actual situations,
the $f^2$-$\Gamma_3$ doublet is mainly composed of two singlets
between $\Gamma_7$ and $\Gamma_8$ orbitals,
and then, we assume an antiferromagnetic interaction
between the $\Gamma_7$ and $\Gamma_8$ orbitals.
Such an interaction would be justified
by perturbatively including the effects of the sixth order terms in the CEF,
which cannot be included as a one-electron potential for $j=5/2$ states
but are indispensable to stabilize the $\Gamma_3$ state.~\cite{Hotta2006}

The model Hamiltonian is
\begin{equation}
  H=H_{\text{kin}}+H_{\text{loc}}.
\end{equation}
$H_{\text{kin}}$ is the kinetic energy term which we will discuss later.
The local part is given by
\begin{equation}
  H_{\text{loc}}
  =
  \Delta \sum_{\bm{r}}(n_{\bm{r}8}-n_{\bm{r}7})
  +J_{78} \sum_{\bm{r}} \bm{s}_{\bm{r}7} \cdot \bm{s}_{\bm{r}8},
\end{equation}
where
\begin{subequations}
\begin{align}
  n_{\bm{r} 7}&=\sum_{\sigma}c^{\dagger}_{\bm{r} 7 \sigma}c_{\bm{r} 7 \sigma},\\
  n_{\bm{r} 8}&=
  \sum_{\tau \sigma}c^{\dagger}_{\bm{r} \tau \sigma}c_{\bm{r} \tau \sigma},\\
  \bm{s}_{\bm{r} 7}&=\frac{1}{2}\sum_{\sigma \sigma'}
  c^{\dagger}_{\bm{r} 7 \sigma} \bm{\sigma}_{\sigma \sigma'} c_{\bm{r} 7 \sigma'},\\
  \bm{s}_{\bm{r} 8}&=\frac{1}{2}\sum_{\tau \sigma \sigma'}
  c^{\dagger}_{\bm{r} \tau \sigma} \bm{\sigma}_{\sigma \sigma'} c_{\bm{r} \tau \sigma'}.
\end{align}
\end{subequations}
Here, $\tau=\alpha$ or $\beta$ and $\bm{\sigma}$ are the Pauli matrices.
$\Delta$ denotes the CEF level splitting [see Fig.~\ref{level_scheme}(a)]
and $J_{78}$ denotes the coupling constant of the antiferromagnetic interaction
between the $\Gamma_7$ and $\Gamma_8$ orbitals [see Fig.~\ref{level_scheme}(b)].

Then, for a sufficiently large $J_{78}$,
the $f^2$ ground states are spin singlets composed
of the $\Gamma_7$ and $\Gamma_8$ orbitals [see Fig.~\ref{level_scheme}(b)]:
\begin{equation}
  \begin{split}
    |\tau (\bm{r}) \rangle
    &\equiv
    \frac{1}{\sqrt{2}}
    (c^{\dagger}_{\bm{r} \tau \uparrow}c^{\dagger}_{\bm{r} 7 \downarrow}
    -c^{\dagger}_{\bm{r} \tau \downarrow}c^{\dagger}_{\bm{r} 7 \uparrow})|0\rangle\\
    &=
    \frac{i}{\sqrt{2}}
    \sigma^y_{\sigma \sigma'}
    c^{\dagger}_{\bm{r} \tau \sigma}c^{\dagger}_{\bm{r} 7 \sigma'}|0\rangle\\
    &\equiv
    B_{\sigma \sigma'}
    c^{\dagger}_{\bm{r} \tau \sigma}c^{\dagger}_{\bm{r} 7 \sigma'}|0\rangle.
  \end{split}
\end{equation}
The repeated indices should be summed hereafter.
These states constitute a basis of the $\Gamma_3$ representation
of cubic symmetry.

Note that the present model is one of the simplest models
to realize the $\Gamma_3$ ground state
and we should improve it if we deal with the CEF excited states.
For example, when we accommodate two electrons in the $\Gamma_8$ orbitals,
we obtain six states with energy $2\Delta$,
but they should split into three levels.
To describe such splitting in the CEF excited states,
it is necessary to include the interactions between $\Gamma_8$ orbitals.
Thus, we should restrict ourselves to low energy states
around the $\Gamma_3$ CEF ground state in the present simplified model.

We consider the exchange process
between nearest-neighbor sites with the $\Gamma_3$ ground state.
Among the intermediate $f^1$-$f^3$ states,
we consider only the lowest energy states.
If the $f^3$ site has zero or two $\Gamma_7$ electrons,
it cannot gain the energy from the antiferromagnetic interaction.
Concerning the $f^1$ states,
we assume that the $\Gamma_7$ state has lower energy, i.e., $\Delta>0$.
However, $\Delta$ should be sufficiently smaller than $J_{78}$
for the realization of the $\Gamma_3$ ground state in the $f^2$ configurations.
Then, each site should have one $\Gamma_7$ electron
in the intermediate states.
That is, only the hopping between the $\Gamma_8$ orbitals is allowed.
In the following, we explicitly write the intermediate states
and evaluate the matrix elements of the exchange processes.

The intermediate $f^1$ states are the $\Gamma_7$ states
[see Fig.~\ref{level_scheme}(a)],
\begin{equation}
  |\sigma(\bm{r}) \rangle
  \equiv
  c^{\dagger}_{\bm{r} 7 \sigma} |0\rangle.
\end{equation}
We calculate
the matrix elements of the annihilation operator of the $\Gamma_8$ electron
between the $f^1$ and the $\Gamma_3$ states.
The effect of the annihilation operator on the $\Gamma_3$ state
is written as
\begin{equation}
  \begin{split}
    c_{\bm{r} \tau \sigma}|\tau'(\bm{r})\rangle
    &=
    c_{\bm{r} \tau \sigma}
    B_{\sigma' \sigma''}
    c^{\dagger}_{\bm{r} \tau' \sigma'}
    c^{\dagger}_{\bm{r} 7 \sigma''}
    |0\rangle\\
    &=
    \delta_{\tau \tau'}
    B_{\sigma \sigma'}
    c^{\dagger}_{\bm{r} 7 \sigma'}
    |0\rangle\\
    &\equiv
    B^{\tau'}_{\tau \sigma; \sigma'}
    |\sigma'(\bm{r}) \rangle.
  \end{split}
\end{equation}
Then, we obtain the matrix element as
\begin{equation}
  \langle \sigma'(\bm{r}) |
  c_{\bm{r} \tau \sigma}|\tau'(\bm{r})\rangle
  =
  B^{\tau'}_{\tau \sigma; \sigma'}.
\end{equation}
Note that
$B^{\tau' *}_{\tau \sigma; \sigma'}=B^{\tau'}_{\tau \sigma; \sigma'}$,
since we have defined these states with real coefficients
from the basis $\Gamma_7$ and $\Gamma_8$ states.

The ground states among the $f^3$ states
for a strong antiferromagnetic interaction $J_{78}$
between the $\Gamma_7$ and $\Gamma_8$ orbitals are
the $\Gamma_6$ states [see Fig.~\ref{level_scheme}(c)],
\begin{subequations}
\begin{align}
  \begin{split}
    |\tilde{\uparrow}(\bm{r}) \rangle
    \equiv&
    \frac{1}{\sqrt{6}}
    (
    2c^{\dagger}_{\bm{r} \alpha \uparrow}
    c^{\dagger}_{\bm{r} \beta \uparrow}
    c^{\dagger}_{\bm{r} 7 \downarrow}\\
    &-c^{\dagger}_{\bm{r} \alpha \uparrow}
    c^{\dagger}_{\bm{r} \beta \downarrow}
    c^{\dagger}_{\bm{r} 7 \uparrow}
    -c^{\dagger}_{\bm{r} \alpha \downarrow}
    c^{\dagger}_{\bm{r} \beta \uparrow}
    c^{\dagger}_{\bm{r} 7 \uparrow}
    )
    |0\rangle\\
    =&
    \frac{1}{\sqrt{3}}
    \left[
      c^{\dagger}_{\bm{r} \alpha \uparrow}|\beta(\bm{r})\rangle
      -c^{\dagger}_{\bm{r} \beta \uparrow}|\alpha(\bm{r})\rangle
    \right],
  \end{split}\\
  \begin{split}
    |\tilde{\downarrow}(\bm{r}) \rangle
    \equiv&
    \frac{1}{\sqrt{6}}
    (
    2c^{\dagger}_{\bm{r} \alpha \downarrow}
    c^{\dagger}_{\bm{r} \beta \downarrow}
    c^{\dagger}_{\bm{r} 7 \uparrow}\\
    &-c^{\dagger}_{\bm{r} \alpha \downarrow}
    c^{\dagger}_{\bm{r} \beta \uparrow}
    c^{\dagger}_{\bm{r} 7 \downarrow}
    -c^{\dagger}_{\bm{r} \alpha \uparrow}
    c^{\dagger}_{\bm{r} \beta \downarrow}
    c^{\dagger}_{\bm{r} 7 \downarrow}
    )
    |0\rangle\\
    =&
    \frac{1}{\sqrt{3}}
    \left[
      -c^{\dagger}_{\bm{r} \alpha \downarrow}|\beta(\bm{r})\rangle
      +c^{\dagger}_{\bm{r} \beta \downarrow}|\alpha(\bm{r})\rangle
    \right].
  \end{split}
\end{align}
\end{subequations}
Note that,
in a local model considering all the 14 $f$-orbitals,
we obtain the $\Gamma_6$ ground state when we accommodate three electrons
for a realistic parameter set to obtain
a $\Gamma_3$ ground state in an $f^2$ case.~\cite{Hotta2006}
Thus, the intermediate $\Gamma_6$ state is reasonable.
Note also that
the states
$c^{\dagger}_{\bm{r} \alpha \sigma}|\beta(\bm{r})\rangle
+c^{\dagger}_{\bm{r} \beta \sigma}|\alpha(\bm{r})\rangle$
are represented by
(spin singlet composed of two $\Gamma_8$ orbitals)$\otimes \Gamma_7$
and they do not gain the antiferromagnetic energy.
The matrix element of the creation operator is given by
\begin{equation}
  \langle \tilde{\sigma}'(\bm{r})|c^{\dagger}_{\bm{r} \tau \sigma}
  | \tau'(\bm{r}) \rangle
  =
  i\sigma^y_{\tau \tau'}
  \sigma^z_{\sigma \sigma'}
  \frac{\sqrt{3}}{2}
  \equiv
  \tilde{B}^{\tau'}_{\tau \sigma; \sigma'}.
\end{equation}
We note that
$\tilde{B}^{\tau' *}_{\tau \sigma; \sigma'}=\tilde{B}^{\tau'}_{\tau \sigma; \sigma'}$.

\section{Hopping}\label{hopping}
The hopping processes are described
by the kinetic energy term of the Hamiltonian for the $\Gamma_8$ orbitals:
\begin{equation}
  \begin{split}
    H_{\text{kin}}
    &=\sum_{\bm{r},\bm{\mu},\tau,\sigma,\tau^{\prime},\sigma^{\prime}}
    c^{\dagger}_{\bm{r}+\bm{\mu} \tau \sigma}
    t^{\bm{\mu}}_{\tau \sigma; \tau^{\prime} \sigma^{\prime}}
    c_{\bm{r} \tau^{\prime} \sigma^{\prime}},\\
    &=\sum_{\bm{r},\bm{\mu},\nu,\nu^{\prime}}
    c^{\dagger}_{\bm{r}+\bm{\mu} \nu}
    t^{\bm{\mu}}_{\nu \nu^{\prime}}
    c_{\bm{r} \nu^{\prime}},
  \end{split}
\end{equation}
where the vector $\bm{\mu}$ connects nearest-neighbor sites.
Here, we have introduced an abbreviation $\nu=(\tau,\sigma)$.
Since $H_{\text{kin}}$ is Hermitian,
$t^{\bm{\mu} *}_{\nu \nu'}=t^{-\bm{\mu}}_{\nu' \nu}$.

In this study,
we consider only the $\sigma$ bonding $(ff\sigma)$
for the hopping integrals.
Although the hopping integrals were derived
in Ref.~\onlinecite{Hotta2003} for the sc lattice
and in Ref.~\onlinecite{Kubo2005PRB} for the other lattices,
here we write down again the hopping integrals for readers' convenience.
To write out the hopping integral $t^{\bm{\mu}}$
for each lattice structure concisely,
we define $4\times4$ matrices as follows
\begin{subequations}
\begin{align}
  \tilde{1}_{\tau \sigma; \tau^{\prime} \sigma^{\prime}}
  &\equiv \delta_{\tau \tau^{\prime}} \delta_{\sigma \sigma^{\prime}},\\
  \tilde{\bm{\tau}}_{\tau \sigma; \tau^{\prime} \sigma^{\prime}}
  &\equiv \bm{\sigma}_{\tau \tau^{\prime}} \delta_{\sigma \sigma^{\prime}},\\
  \tilde{\bm{\sigma}}_{\tau \sigma; \tau^{\prime} \sigma^{\prime}}
  &\equiv \delta_{\tau \tau^{\prime}} \bm{\sigma}_{\sigma \sigma^{\prime}},\\
  \tilde{\eta}^{\pm}
  &\equiv (\pm \sqrt{3}\tilde{\tau}^x-\tilde{\tau}^z)/2.
\end{align}
\end{subequations}
Then, the hopping integrals for the sc lattice are given by
\begin{subequations}
\begin{align}
  t^{(1,0,0)} &= [\tilde{1}-\tilde{\eta}^+]t_1,\\
  t^{(0,1,0)} &= [\tilde{1}-\tilde{\eta}^-]t_1,\\
  t^{(0,0,1)} &= [\tilde{1}-\tilde{\tau}^z]t_1,
\end{align}
\end{subequations}
where we have set the lattice constant as unity
and $t_1$=$3(ff\sigma)/14$.
For the bcc lattice,
\begin{subequations}
\begin{align}
  t^{(1/2,1/2,1/2)} &= [\tilde{1}
    +\tilde{\tau}^y
    (+\tilde{\sigma}^x+\tilde{\sigma}^y+\tilde{\sigma}^z)/\sqrt{3}]t_2,
  \label{t111}\\
  t^{(-1/2,1/2,1/2)} &= [\tilde{1}                                     
    +\tilde{\tau}^y
    (+\tilde{\sigma}^x-\tilde{\sigma}^y-\tilde{\sigma}^z)/\sqrt{3}]t_2,\\
  t^{(1/2,-1/2,1/2)} &= [\tilde{1}                                     
    +\tilde{\tau}^y
    (-\tilde{\sigma}^x+\tilde{\sigma}^y-\tilde{\sigma}^z)/\sqrt{3}]t_2,\\
  t^{(1/2,1/2,-1/2)} &= [\tilde{1}                                     
    +\tilde{\tau}^y
    (-\tilde{\sigma}^x-\tilde{\sigma}^y+\tilde{\sigma}^z)/\sqrt{3}]t_2,
  \label{t11-1}
\end{align}
\end{subequations}
with $t_2$=$2(ff\sigma)/21$.
For the fcc lattice,
\begin{subequations}
\begin{align}
t^{(0,1/2,1/2)}&=
[\tilde{1}+(\tilde{\eta}^+ -4\sqrt{3} \tilde{\tau}^y
\tilde{\sigma}^x)/7]t_3, \label{t011}\\
t^{(1/2,0,1/2)}&=                                                  
[\tilde{1}+(\tilde{\eta}^- -4\sqrt{3} \tilde{\tau}^y
\tilde{\sigma}^y)/7]t_3, \\
t^{(1/2,1/2,0)}&=                                                  
[\tilde{1}+(\tilde{\tau}^z -4\sqrt{3} \tilde{\tau}^y
\tilde{\sigma}^z)/7]t_3, \\
t^{(0,1/2,-1/2)}&=                                                 
[\tilde{1}+(\tilde{\eta}^+ +4\sqrt{3} \tilde{\tau}^y
\tilde{\sigma}^x)/7]t_3, \\
t^{(-1/2,0,1/2)}&=                                                 
[\tilde{1}+(\tilde{\eta}^- +4\sqrt{3} \tilde{\tau}^y
\tilde{\sigma}^y)/7]t_3, \\
t^{(1/2,-1/2,0)}&=                                                 
[\tilde{1}+(\tilde{\tau}^z +4\sqrt{3} \tilde{\tau}^y
\tilde{\sigma}^z)/7]t_3,
\end{align}
\end{subequations}
with $t_3$=$(ff\sigma)/8$.
Except for the sc lattice,
the hopping integrals are complex numbers and dependent on $\sigma$.
Note that $t^{-\bm{\mu}}=t^{\bm{\mu}}$.

\section{Multipole interaction}\label{results}
By employing the second-order perturbation theory
with respect to $H_{\text{kin}}$,
we derive the effective Hamiltonian:
\begin{equation}
  \begin{split}
    H^{\text{(eff)}}
    = \sum_{a,b,u}\sum_{m \ne 0}
    |0, a \rangle \langle 0, a| &H_{\text{kin}}
    \frac{|m, u \rangle \langle m, u|}{E_0-E_m}\\
    \times
    &H_{\text{kin}} |0, b \rangle \langle 0, b|.
  \end{split}
\end{equation}
Here, $|0, a \rangle$ is a ground state
without $H_{\text{kin}}$ with the energy $E_0$
and $|m, u \rangle$ is an $m$-th excited state
with the energy $E_m$.
In the following, we consider only the first excited states
among the intermediate states,
which are described by a pair of nearest-neighboring
$f^1$ and $f^3$ sites discussed above.
Then, we need to evaluate the following matrix element:
\begin{equation}
  \begin{split}
    -&\Delta E
    \times H^{\text{(eff)}}_{\tau_1 \tau_2; \tau'_1 \tau'_2}(\bm{r}_1,\bm{r}_2)\\
    =
    \sum_{u}
    &\langle
    \tau_1(\bm{r}_1) \
    \tau_2(\bm{r}_2)
    |H_{\text{kin}}
    |1,u \rangle \\
    \times&\langle 1,u|
    H_{\text{kin}}|
    \tau'_1(\bm{r}_1) \
    \tau'_2(\bm{r}_2)
    \rangle,
  \end{split}
\end{equation}
where $\Delta E=E_1-E_0=J_{78}/2$.
This matrix element denotes the transitions of the states:
$\tau'_1 \rightarrow \tau_1$ at $\bm{r}_1$
and
$\tau'_2 \rightarrow \tau_2$ at $\bm{r}_2$.
The part of the element in which
the intermediate $f^1$ state is located at $\bm{r}_2$
and $f^3$ state is located at $\bm{r}_1$ is given by
\begin{equation}
  \begin{split}
    & \sum_u \langle
    \tau_1(\bm{r}_1) \
    \tau_2(\bm{r}_2)
    |
    c^{\dagger}_{\bm{r}_2 \nu_2}
    t^{\bm{r}_2-\bm{r}_1}_{\nu_2 \nu_1}
    c_{\bm{r}_1 \nu_1}
    |1, u \rangle\\
    &\times
    \langle 1, u|
    c^{\dagger}_{\bm{r}_1 \nu'_1}
    t^{\bm{r}_1-\bm{r}_2}_{\nu'_1 \nu'_2}
    c_{\bm{r}_2 \nu'_2}
    |
    \tau'_1(\bm{r}_1) \
    \tau'_2(\bm{r}_2)
    \rangle\\
    =&
    t^{\bm{r}_2-\bm{r}_1}_{\nu_2 \nu_1}
    t^{\bm{r}_1-\bm{r}_2}_{\nu'_1 \nu'_2}\\
    &\times
    \langle
    \tau_1(\bm{r}_1)
    |
    c_{\bm{r}_1 \nu_1}
    |\tilde{\sigma}_1(\bm{r}_1)\rangle\langle\tilde{\sigma}_1(\bm{r}_1)|
    c^{\dagger}_{\bm{r}_1 \nu'_1}
    |
    \tau'_1(\bm{r}_1)
    \rangle\\
    &\times
    \langle
    \tau_2(\bm{r}_2)
    |
    c^{\dagger}_{\bm{r}_2 \nu_2}
    |\sigma_2(\bm{r}_2)\rangle\langle\sigma_2(\bm{r}_2)|
    c_{\bm{r}_2 \nu'_2}
    |
    \tau'_2(\bm{r}_2)
    \rangle\\
    =&
    t^{\bm{r}_2-\bm{r}_1}_{\nu_2 \nu_1}
    t^{\bm{r}_1-\bm{r}_2}_{\nu'_1 \nu'_2}
    \tilde{B}^{\tau_1 *}_{\nu_1 \sigma_1}
    \tilde{B}^{\tau'_1}_{\nu'_1 \sigma_1}
    B^{\tau_2 *}_{\nu_2 \sigma_2}
    B^{\tau'_2}_{\nu'_2 \sigma_2}\\
    =&
    \text{Tr}[
      B^{\tau_2 \text{T}}
      t^{\bm{r}_2-\bm{r}_1}
      \tilde{B}^{\tau_1}
      \tilde{B}^{\tau'_1 \text{T}}
      t^{\bm{r}_1-\bm{r}_2}
      B^{\tau'_2}
    ],
  \end{split}
\end{equation}
where Tr and T denote trace and transpose of a matrix, respectively.
Similarly, the part of the element with
the intermediate $f^1$ state at $\bm{r}_1$ and $f^3$ state at $\bm{r}_2$
is given by
\begin{equation}
  \begin{split}
    & \sum_u \langle
    \tau_1(\bm{r}_1) \
    \tau_2(\bm{r}_2)
    |
    c^{\dagger}_{\bm{r}_1 \nu'_1}
    t^{\bm{r}_1-\bm{r}_2}_{\nu'_1 \nu'_2}
    c_{\bm{r}_2 \nu'_2}
    |1,u \rangle\\
    &\times
    \langle 1,u|
    c^{\dagger}_{\bm{r}_2 \nu_2}
    t^{\bm{r}_2-\bm{r}_1}_{\nu_2 \nu_1}
    c_{\bm{r}_1 \nu_1}
    |
    \tau'_1(\bm{r}_1) \
    \tau'_2(\bm{r}_2)
    \rangle\\
    =&
    t^{\bm{r}_2-\bm{r}_1}_{\nu_2 \nu_1}
    t^{\bm{r}_1-\bm{r}_2}_{\nu'_1 \nu'_2}\\
    &\times\langle
    \tau_1(\bm{r}_1)
    |
    c^{\dagger}_{\bm{r}_1 \nu'_1}
    |\sigma_1(\bm{r}_1)\rangle\langle\sigma_1(\bm{r}_1)|
    c_{\bm{r}_1 \nu_1}
    |
    \tau'_1(\bm{r}_1)
    \rangle\\
    &\times
    \langle
    \tau_2(\bm{r}_2)
    |
    c_{\bm{r}_2 \nu'_2}
    |\tilde{\sigma}_2(\bm{r}_2)\rangle\langle\tilde{\sigma}_2(\bm{r}_2)|
    c^{\dagger}_{\bm{r}_2 \nu_2}
    |
    \tau'_2(\bm{r}_2)
    \rangle\\
    =&
    t^{\bm{r}_2-\bm{r}_1}_{\nu_2 \nu_1}
    t^{\bm{r}_1-\bm{r}_2}_{\nu'_1 \nu'_2}
    B^{\tau_1 *}_{\nu'_1 \sigma_1}
    B^{\tau'_1}_{\nu_1 \sigma_1}
    \tilde{B}^{\tau_2 *}_{\nu'_2 \sigma_2}
    \tilde{B}^{\tau'_2}_{\nu_2 \sigma_2}\\
    =&
    \text{Tr}[
      \tilde{B}^{\tau'_2 \text{T}}
      t^{\bm{r}_2-\bm{r}_1}
      B^{\tau'_1}
      B^{\tau_1 \text{T}}
      t^{\bm{r}_1-\bm{r}_2}
      \tilde{B}^{\tau_2}
    ]\\
    =&
    \text{Tr}[
      \tilde{B}^{\tau_2 \text{T}}
      t^{\bm{r}_2-\bm{r}_1 *}
      B^{\tau_1}
      B^{\tau'_1 \text{T}}
      t^{\bm{r}_1-\bm{r}_2 *}
      \tilde{B}^{\tau'_2}
    ].
  \end{split}
\end{equation}
Then, the total matrix element of the effective Hamiltonian
is
\begin{equation}
  \begin{split}
    &-\Delta E
    \times H^{\text{(eff)}}_{\tau_1 \tau_2; \tau'_1 \tau'_2}(\bm{r}_1,\bm{r}_2)\\
    =&
    \text{Tr}[
      B^{\tau_2 \text{T}}
      t^{\bm{r}_2-\bm{r}_1}
      \tilde{B}^{\tau_1}
      \tilde{B}^{\tau'_1 \text{T}}
      t^{\bm{r}_1-\bm{r}_2}
      B^{\tau'_2}
    ]\\
    &+\text{Tr}[
      \tilde{B}^{\tau_2 \text{T}}
      t^{\bm{r}_2-\bm{r}_1 *}
      B^{\tau_1}
      B^{\tau'_1 \text{T}}
      t^{\bm{r}_1-\bm{r}_2 *}
      \tilde{B}^{\tau'_2}
    ].
  \end{split}
\end{equation}
By straightforward algebraic calculations of the matrices
$B^{\tau}$,  $\tilde{B}^{\tau}$, and $t^{\bm{\mu}}$,
we can evaluate this equation for each lattice structure.

The obtained effective Hamiltonian can be rewritten
by using the multipole operators
for the $\Gamma_3$ state
defined by
\begin{subequations}
\begin{align}
  O^0_{2 \bm{r}}
  &=\sum_{\tau \tau'} |\tau(\bm{r})\rangle \sigma^z_{\tau \tau'}
  \langle \tau'(\bm{r})|,\\
  O^2_{2 \bm{r}}
  &=\sum_{\tau \tau'} |\tau(\bm{r})\rangle \sigma^x_{\tau \tau'}
  \langle \tau'(\bm{r})|,\\
  T_{xyz \bm{r}}
  &=\sum_{\tau \tau'} |\tau(\bm{r})\rangle \sigma^y_{\tau \tau'}
  \langle \tau'(\bm{r})|
  \label{Txyz}.
\end{align}
\end{subequations}
$O^0_{2 \bm{r}}$ and $O^2_{2 \bm{r}}$ are the quadrupole moments
with $\Gamma_{3g}$ symmetry
and
$T_{xyz \bm{r}}$ is the octupole moment
with $\Gamma_{2u}$ symmetry.

In the following,
we show the derived multipole interactions
in the Fourier transformed from.
Previously,
we have also derived the multipole interactions for $f^1$ systems
with the $\Gamma_8$ CEF ground state
by a similar method.~\cite{Kubo2005PRBR,Kubo2005PRB}
We will compare the multipole interactions for
the present $f^2$-$\Gamma_3$ model
with those for the $f^1$-$\Gamma_8$ model.

\subsection{sc lattice}
\begin{figure}
  \includegraphics[width=0.85\linewidth]
  {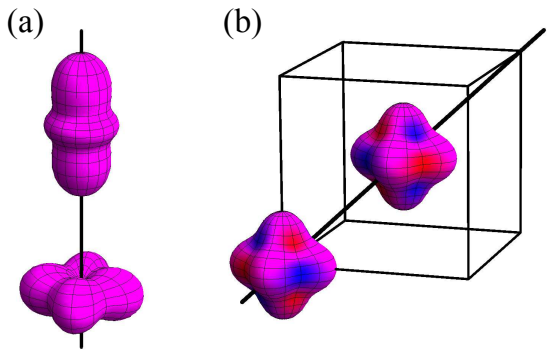}
  \caption{\label{nearest_neighbor_interaction}
    (Color online)
    Schematic figures of
    the electronic states on the nearest-neighboring sites
    preferred by the interaction
    (a) along the $z$ direction
    (antiferro arrangement of the $O^0_2$ moments)
    and
    (b) along [111] direction
    (antiferro arrangement of the $T_{xyz}$ moments).
    The gradation of color in (b) indicates
    the anisotropic distribution of the dipole moment.
  }
\end{figure}

For the sc lattice,
we obtain only the following quadrupole interaction,
\begin{equation}
  \begin{split}
    H^{\text{(eff)}}&=
    \frac{3}{2}\sum_{\bm{q}}
    \biggl[
      \cos q_z O^0_{2 \bm{q}}O^0_{2 -\bm{q}}\\
      +&\cos q_x \frac{1}{4}
      (\sqrt{3}O^2_{2 \bm{q}}-O^0_{2 \bm{q}})
      (\sqrt{3}O^2_{2 -\bm{q}}-O^0_{2 -\bm{q}})\\
      +&\cos q_y \frac{1}{4}
      (\sqrt{3}O^2_{2 \bm{q}}+O^0_{2 \bm{q}})
      (\sqrt{3}O^2_{2 -\bm{q}}+O^0_{2 -\bm{q}})
    \biggr],
  \end{split}
\end{equation}
in the unit of $t^2_1/\Delta E$.
We can intuitively understand why this interaction is dominant
since the $z$ direction is congenial to $3z^2-r^2$ ($O^0_2$) symmetry
[see Fig.~\ref{nearest_neighbor_interaction}(a)].
Also in the $\Gamma_8$ model,
this quadrupole interaction is the main interaction
and the $\Gamma_{2u}$ octupole interaction is absent.

Note that the derived model is the same as
a model for ferromagnetic insulating manganites
describing only the orbital degrees of freedom
of $e_g$ electrons,~\cite{Shiina1997,Brink1999,Ishihara2000,Kubo2002JPSJNo5}
except for the overall coefficient.
This model has continuously degenerate ground states in the mean-field level
due to the frustration which originates from the anisotropic interaction.

If we approximate the ordering vector for PrPb$_3$ by $\bm{q}=(\pi,\pi,0)$
and assume this ordering vector to the model,
we obtain an ordering of the $O^2_2$ moment by the mean-field theory.
This is out of accord with the experimental indications
of the $O^0_2$ ordering.~\cite{Onimaru2004,Onimaru2005}
For the $O^0_2$ ordering in PrPb$_3$,
we need to improve the present theory,
for example,
by considering the long-range interactions, which are also important
to stabilize the incommensurate ordering observed in PrPb$_3$.

\subsection{bcc lattice}
For the bcc lattice, we obtain only the following octupole interaction,
\begin{equation}
  \begin{split}
    H^{\text{(eff)}}
    =
    6\sum_{\bm{q}}
    &\cos(q_x/2)\cos(q_y/2)\cos(q_z/2)\\
    &\times T_{xyz \bm{q}}T_{xyz -\bm{q}},
  \end{split}
\end{equation}
in the unit of $t^2_2/\Delta E$
and the ground state of this effective model is the staggered ordered state
of the octupole moments.
Since the [111] direction is congenial to $xyz$ symmetry,
we can naturally understand that this interaction is dominant
[see Fig.~\ref{nearest_neighbor_interaction}(b)].
Also in the $\Gamma_8$ model,
this octupole interaction is the main interaction
and the $\Gamma_{3g}$ quadrupole interaction is absent.
The existence of the $\tilde{\tau}_y$ term in Eqs.~\eqref{t111}--\eqref{t11-1}
suggests the interaction of the $T_{xyz}$ moments [Eq.~\eqref{Txyz}],
in accord with the present result.
We will discuss this point in the next subsection.

If ordering of this type of octupole moments occurs,
we will observe an anomaly in the specific heat
as in an ordinary phase transition,
but the determination of the order parameter will be challenging
since neither the dipole nor quadrupole moments will be induced,
in contrast to the octupole order in NpO$_2$ and in Ce$_x$La$_{1-x}$B$_6$
where quadrupole moments are induced.~\cite{Paixao2002,Akatsu2003,
  Kubo2003,Kubo2004,Tokunaga2005,Kubo2005PRBR,Kubo2005PRB,Kubo2005PRBB,
  Inami2014}

The possibility of the ordering of this octupole moment had also been discussed
for an $e_g$-electron model for manganites
in a ferromagnetic metallic phase.~\cite{Takahashi2000,Maezono2000,
  Brink2001,Khomskii2001}
However, it was revealed that this ordering is unstable
against fluctuations beyond the mean-field theory.~\cite{Kubo2002JPSJNo1}
On the other hand, in the present model for $f$ electrons,
we have a clear picture for the realization of the $T_{xyz}$ ordering
and it should be stable against fluctuations.

Note also that, in the diamond structure,
the nearest-neighbor sites locate $(1/4,1/4,1/4)$
and so on from the origin,
and thus, we obtain only the $\Gamma_{2u}$ octupole interaction
as in the bcc lattice.
Therefore, we may expect strong fluctuations of the octupole moments
in the Pr 1-2-20 systems, in which Pr ions form the diamond structure.

\subsection{fcc lattice}
For the fcc lattice,
we obtain both quadrupole and octupole interactions,
\begin{equation}
  \begin{split}
    H^{\text{(eff)}}
    =
    \frac{3}{49}&\sum_{\bm{q}}
    \biggl[
      \cos(q_x/2)\cos(q_y/2) O^0_{2 \bm{q}}O^0_{2 -\bm{q}}\\
      +&\cos(q_y/2)\cos(q_z/2)\\
      \times&
      \frac{1}{4}
      (\sqrt{3}O^2_{2 \bm{q}}-O^0_{2 \bm{q}})
      (\sqrt{3}O^2_{2 -\bm{q}}-O^0_{2 -\bm{q}})\\
      +&\cos(q_z/2)\cos(q_x/2)\\
      \times&
      \frac{1}{4}
      (\sqrt{3}O^2_{2 \bm{q}}+O^0_{2 \bm{q}})
      (\sqrt{3}O^2_{2 -\bm{q}}+O^0_{2 -\bm{q}})
    \biggr]\\
    +\frac{144}{49}&\sum_{\bm{q}}
    [\cos(q_x/2)\cos(q_y/2)\\
      +&\cos(q_y/2)\cos(q_z/2)\\
      +&\cos(q_z/2)\cos(q_x/2)]
    T_{xyz \bm{q}}T_{xyz -\bm{q}},
  \end{split}
\end{equation}
in the unit of $t^2_3/\Delta E$.
Broadly speaking,
the fcc lattice has characteristics between the sc and bcc lattices
and, as a result, we have obtained both quadrupole and octupole interactions.
In the $\Gamma_8$ model,
the $\Gamma_{2u}$ octupole interaction competes
with a $\Gamma_{4u}$ dipole and octupole interaction
and a $\Gamma_{5u}$ octupole interaction,
which are absent here since the $\Gamma_3$ doublet
does not have these degrees of freedom.
The $\Gamma_{3g}$ quadruple interaction is weak but finite
in the $\Gamma_8$ model
and it is also similar to the present $\Gamma_3$ model.
Since the octupole interaction is larger than the quadrupole interaction,
the ground state of the model is the staggered ordered state
of the octupole moments
at least in the mean-field theory within two-sublattice structures.

In general, the quadruple and octupole interactions
may compete with each other, but at least in the present simple model,
the octupole interaction is dominant.
The large difference in the magnitude of the interactions
originates from the coefficients in the hopping integral.
The ratio of the coefficient of $\tilde{\eta}^{+}$
to that of $\tilde{\tau}^y$ is 1 to $-4\sqrt{3}$ in Eq.~\eqref{t011}
and the ratio of the square of them is 1 to 48;
it is the ratio of the quadrupole and octupole interactions.

In the sc lattice,
the hopping integral does not have a $\tilde{\tau}^y$ term
and the octupole interaction is absent.
In the bcc lattice,
the hopping integral does not have an $\tilde{\eta}^+$, $\tilde{\eta}^-$
or $\tilde{\tau}^z$ term
and the quadrupole interaction is absent.
However, in general, it is not so simple.
For example, if the hopping is isotropic,
that is, there is no 
$\tilde{\eta}^+$, $\tilde{\eta}^-$, $\tilde{\tau}^z$,
or $\tilde{\tau}^y$ term,
we obtain an isotropic Heisenberg-type interaction,
i.e., both quadrupole and octupole interactions.

\begin{table}[t]
  \caption{\label{dominant_interactions}
    Dominant interactions in each lattice
    for the $f^2$-$\Gamma_3$ model (present study)
    and for the $f^1$-$\Gamma_8$ model
    (Refs.~\onlinecite{Kubo2005PRBR,Kubo2005PRB}).
  }
  \begin{ruledtabular}
     \begin{tabular}{cccc}
       CEF state & sc & bcc & fcc \\
       \hline
       $f^2$-$\Gamma_3$ & $\Gamma_{3g}$ quadrupole & $\Gamma_{2u}$ octupole
       & $\Gamma_{2u}$ octupole\\
       $f^1$-$\Gamma_8$ & $\Gamma_{3g}$ quadrupole & $\Gamma_{2u}$ octupole
       & $\Gamma_{2u}$, $\Gamma_{4u}$, $\Gamma_{5u}$
     \end{tabular}
  \end{ruledtabular}
\end{table}
In Table~\ref{dominant_interactions},
we summarize the dominant interactions in each lattice
for the $f^2$-$\Gamma_3$ model obtained here 
and for the the $f^1$-$\Gamma_8$ model
(Refs.~\onlinecite{Kubo2005PRBR,Kubo2005PRB}).

\section{Multipole interactions in another simplified model}
In this section, we discuss another simple model to describe
the $\Gamma_3$ CEF ground state.
Here, we omit the $\Gamma_7$ orbital and construct
the $\Gamma_3$ states only from the $\Gamma_8$ orbitals.
This model is too simple to discuss realistic situations,
but by comparing with the results in the previous section,
we can recognize how much the multipole interactions are altered
by the choice of the model.
By omitting the $\Gamma_7$ orbital, the derivation
of the multipole interactions becomes rather simple
since the intermediate $f^1$-$f^3$ states do not split.

For the sc lattice, we obtain no multipole interaction, i.e.,
the second-order perturbation theory merely gives an energy shift.
It means that this model is too simple.
For the bcc and fcc lattices, we obtain only the octupole interaction.
Thus, the dominance of the octupole interaction in the bcc and fcc lattices
is common between the models in this section and in the previous sections.

Therefore, we expect that the characteristic features
of the multipole interactions
summarized in Table~\ref{dominant_interactions} will not change,
even if we use different ways to construct the $\Gamma_3$ state,
except for special cases such as the sc lattice in this section.

\section{Summary}
We have investigated the multipole interactions
by the second-order perturbation theory
to a simple model for the $f^2$ ions with the $\Gamma_3$ non-Kramers doublet
ground state under a cubic CEF,
in particular, by paying attention to the lattice structure.
We have obtained the $\Gamma_{3g}$ quadrupole interaction for a sc lattice
and the $\Gamma_{2u}$ octupole interaction for a bcc lattice.
For an fcc lattice, we have obtained both interactions.
These characteristics are the same as those for the $f^1$-$\Gamma_8$ model.
Thus, we expect that such tendencies or correspondences
between the dominant multipole interactions and the lattice structures
are common as long as the ground CEF state
has these multipole degrees of freedom.

While several kinds of multipole order are possible to occur in general,
the $\Gamma_{2u}$ octupole order is particularly fascinating
since it will induce neither the dipole nor quadrupole moments,
even though the specific heat will show an anomaly at the transition point
as in an ordinary phase transition.
In this regard,
it would be interesting to search bcc lattices and diamond structure
for the $\Gamma_{2u}$ order
since we have obtained a strong interaction for this kind of moments
both in the $f^2$-$\Gamma_3$ and $f^1$-$\Gamma_8$ models.

The general forms of the multipole interactions
have been derived in Ref.~\onlinecite{Sakai2003}.
For example, another form of the quadrupole interaction
is possible for a sc lattice in general.
We expect that such components appear when we introduce
hopping integrals other than $(ff\sigma)$.
Thus, we should note that the applicability of the present results
are limited to the cases where the (effective) hopping processes are mainly
described by $(ff\sigma)$.

\begin{acknowledgments}
This work was supported by JSPS KAKENHI Grant Numbers
JP15K05191, 
JP16H04017, 
and JP16H01079 (J-Physics). 
\end{acknowledgments}


\begin{thebibliography}{32}%
\makeatletter
\providecommand \@ifxundefined [1]{%
 \@ifx{#1\undefined}
}%
\providecommand \@ifnum [1]{%
 \ifnum #1\expandafter \@firstoftwo
 \else \expandafter \@secondoftwo
 \fi
}%
\providecommand \@ifx [1]{%
 \ifx #1\expandafter \@firstoftwo
 \else \expandafter \@secondoftwo
 \fi
}%
\providecommand \natexlab [1]{#1}%
\providecommand \enquote  [1]{``#1''}%
\providecommand \bibnamefont  [1]{#1}%
\providecommand \bibfnamefont [1]{#1}%
\providecommand \citenamefont [1]{#1}%
\providecommand \href@noop [0]{\@secondoftwo}%
\providecommand \href [0]{\begingroup \@sanitize@url \@href}%
\providecommand \@href[1]{\@@startlink{#1}\@@href}%
\providecommand \@@href[1]{\endgroup#1\@@endlink}%
\providecommand \@sanitize@url [0]{\catcode `\\12\catcode `\$12\catcode
  `\&12\catcode `\#12\catcode `\^12\catcode `\_12\catcode `\%12\relax}%
\providecommand \@@startlink[1]{}%
\providecommand \@@endlink[0]{}%
\providecommand \url  [0]{\begingroup\@sanitize@url \@url }%
\providecommand \@url [1]{\endgroup\@href {#1}{\urlprefix }}%
\providecommand \urlprefix  [0]{URL }%
\providecommand \Eprint [0]{\href }%
\providecommand \doibase [0]{http://dx.doi.org/}%
\providecommand \selectlanguage [0]{\@gobble}%
\providecommand \bibinfo  [0]{\@secondoftwo}%
\providecommand \bibfield  [0]{\@secondoftwo}%
\providecommand \translation [1]{[#1]}%
\providecommand \BibitemOpen [0]{}%
\providecommand \bibitemStop [0]{}%
\providecommand \bibitemNoStop [0]{.\EOS\space}%
\providecommand \EOS [0]{\spacefactor3000\relax}%
\providecommand \BibitemShut  [1]{\csname bibitem#1\endcsname}%
\let\auto@bib@innerbib\@empty
\bibitem [{\citenamefont {Paix{\~a}o}\ \emph {et~al.}(2002)\citenamefont
  {Paix{\~a}o}, \citenamefont {Detlefs}, \citenamefont {Longfield},
  \citenamefont {Caciuffo}, \citenamefont {Santini}, \citenamefont {Bernhoeft},
  \citenamefont {Rebizant},\ and\ \citenamefont {Lander}}]{Paixao2002}%
  \BibitemOpen
  \bibfield  {author} {\bibinfo {author} {\bibfnamefont {J.~A.}\ \bibnamefont
  {Paix{\~a}o}}, \bibinfo {author} {\bibfnamefont {C.}~\bibnamefont {Detlefs}},
  \bibinfo {author} {\bibfnamefont {M.~J.}\ \bibnamefont {Longfield}}, \bibinfo
  {author} {\bibfnamefont {R.}~\bibnamefont {Caciuffo}}, \bibinfo {author}
  {\bibfnamefont {P.}~\bibnamefont {Santini}}, \bibinfo {author} {\bibfnamefont
  {N.}~\bibnamefont {Bernhoeft}}, \bibinfo {author} {\bibfnamefont
  {J.}~\bibnamefont {Rebizant}}, \ and\ \bibinfo {author} {\bibfnamefont
  {G.~H.}\ \bibnamefont {Lander}},\ }\href@noop {} {\bibfield  {journal}
  {\bibinfo  {journal} {Phys. Rev. Lett.}\ }\textbf {\bibinfo {volume} {89}},\
  \bibinfo {pages} {187202} (\bibinfo {year} {2002})}\BibitemShut {NoStop}%
\bibitem [{\citenamefont {Tokunaga}\ \emph {et~al.}(2005)\citenamefont
  {Tokunaga}, \citenamefont {Homma}, \citenamefont {Kambe}, \citenamefont
  {Aoki}, \citenamefont {Sakai}, \citenamefont {Yamamoto}, \citenamefont
  {Nakamura}, \citenamefont {Shiokawa}, \citenamefont {Walstedt},\ and\
  \citenamefont {Yasuoka}}]{Tokunaga2005}%
  \BibitemOpen
  \bibfield  {author} {\bibinfo {author} {\bibfnamefont {Y.}~\bibnamefont
  {Tokunaga}}, \bibinfo {author} {\bibfnamefont {Y.}~\bibnamefont {Homma}},
  \bibinfo {author} {\bibfnamefont {S.}~\bibnamefont {Kambe}}, \bibinfo
  {author} {\bibfnamefont {D.}~\bibnamefont {Aoki}}, \bibinfo {author}
  {\bibfnamefont {H.}~\bibnamefont {Sakai}}, \bibinfo {author} {\bibfnamefont
  {E.}~\bibnamefont {Yamamoto}}, \bibinfo {author} {\bibfnamefont
  {A.}~\bibnamefont {Nakamura}}, \bibinfo {author} {\bibfnamefont
  {Y.}~\bibnamefont {Shiokawa}}, \bibinfo {author} {\bibfnamefont {R.~E.}\
  \bibnamefont {Walstedt}}, \ and\ \bibinfo {author} {\bibfnamefont
  {H.}~\bibnamefont {Yasuoka}},\ }\href@noop {} {\bibfield  {journal} {\bibinfo
   {journal} {Phys. Rev. Lett.}\ }\textbf {\bibinfo {volume} {94}},\ \bibinfo
  {pages} {137209} (\bibinfo {year} {2005})}\BibitemShut {NoStop}%
\bibitem [{\citenamefont {Kubo}\ and\ \citenamefont
  {Hotta}(2005{\natexlab{a}})}]{Kubo2005PRBR}%
  \BibitemOpen
  \bibfield  {author} {\bibinfo {author} {\bibfnamefont {K.}~\bibnamefont
  {Kubo}}\ and\ \bibinfo {author} {\bibfnamefont {T.}~\bibnamefont {Hotta}},\
  }\href@noop {} {\bibfield  {journal} {\bibinfo  {journal} {Phys. Rev. B}\
  }\textbf {\bibinfo {volume} {71}},\ \bibinfo {pages} {140404(R)} (\bibinfo
  {year} {2005}{\natexlab{a}})}\BibitemShut {NoStop}%
\bibitem [{\citenamefont {Kubo}\ and\ \citenamefont
  {Hotta}(2005{\natexlab{b}})}]{Kubo2005PRB}%
  \BibitemOpen
  \bibfield  {author} {\bibinfo {author} {\bibfnamefont {K.}~\bibnamefont
  {Kubo}}\ and\ \bibinfo {author} {\bibfnamefont {T.}~\bibnamefont {Hotta}},\
  }\href@noop {} {\bibfield  {journal} {\bibinfo  {journal} {Phys. Rev. B}\
  }\textbf {\bibinfo {volume} {72}},\ \bibinfo {pages} {144401} (\bibinfo
  {year} {2005}{\natexlab{b}})}\BibitemShut {NoStop}%
\bibitem [{\citenamefont {Kubo}\ and\ \citenamefont
  {Hotta}(2005{\natexlab{c}})}]{Kubo2005PRBB}%
  \BibitemOpen
  \bibfield  {author} {\bibinfo {author} {\bibfnamefont {K.}~\bibnamefont
  {Kubo}}\ and\ \bibinfo {author} {\bibfnamefont {T.}~\bibnamefont {Hotta}},\
  }\href@noop {} {\bibfield  {journal} {\bibinfo  {journal} {Phys. Rev. B}\
  }\textbf {\bibinfo {volume} {72}},\ \bibinfo {pages} {132411} (\bibinfo
  {year} {2005}{\natexlab{c}})}\BibitemShut {NoStop}%
\bibitem [{\citenamefont {Akatsu}\ \emph {et~al.}(2003)\citenamefont {Akatsu},
  \citenamefont {Goto}, \citenamefont {Nemoto}, \citenamefont {Suzuki},
  \citenamefont {Nakamura},\ and\ \citenamefont {Kunii}}]{Akatsu2003}%
  \BibitemOpen
  \bibfield  {author} {\bibinfo {author} {\bibfnamefont {M.}~\bibnamefont
  {Akatsu}}, \bibinfo {author} {\bibfnamefont {T.}~\bibnamefont {Goto}},
  \bibinfo {author} {\bibfnamefont {Y.}~\bibnamefont {Nemoto}}, \bibinfo
  {author} {\bibfnamefont {O.}~\bibnamefont {Suzuki}}, \bibinfo {author}
  {\bibfnamefont {S.}~\bibnamefont {Nakamura}}, \ and\ \bibinfo {author}
  {\bibfnamefont {S.}~\bibnamefont {Kunii}},\ }\href@noop {} {\bibfield
  {journal} {\bibinfo  {journal} {J. Phys. Soc. Jpn.}\ }\textbf {\bibinfo
  {volume} {72}},\ \bibinfo {pages} {205} (\bibinfo {year} {2003})}\BibitemShut
  {NoStop}%
\bibitem [{\citenamefont {Kubo}\ and\ \citenamefont
  {Kuramoto}(2003)}]{Kubo2003}%
  \BibitemOpen
  \bibfield  {author} {\bibinfo {author} {\bibfnamefont {K.}~\bibnamefont
  {Kubo}}\ and\ \bibinfo {author} {\bibfnamefont {Y.}~\bibnamefont
  {Kuramoto}},\ }\href@noop {} {\bibfield  {journal} {\bibinfo  {journal} {J.
  Phys. Soc. Jpn.}\ }\textbf {\bibinfo {volume} {72}},\ \bibinfo {pages} {1859}
  (\bibinfo {year} {2003})}\BibitemShut {NoStop}%
\bibitem [{\citenamefont {Kubo}\ and\ \citenamefont
  {Kuramoto}(2004)}]{Kubo2004}%
  \BibitemOpen
  \bibfield  {author} {\bibinfo {author} {\bibfnamefont {K.}~\bibnamefont
  {Kubo}}\ and\ \bibinfo {author} {\bibfnamefont {Y.}~\bibnamefont
  {Kuramoto}},\ }\href@noop {} {\bibfield  {journal} {\bibinfo  {journal} {J.
  Phys. Soc. Jpn.}\ }\textbf {\bibinfo {volume} {73}},\ \bibinfo {pages} {216}
  (\bibinfo {year} {2004})}\BibitemShut {NoStop}%
\bibitem [{\citenamefont {Morie}\ \emph {et~al.}(2004)\citenamefont {Morie},
  \citenamefont {Sakakibara}, \citenamefont {Tayama},\ and\ \citenamefont
  {Kunii}}]{Morie2004}%
  \BibitemOpen
  \bibfield  {author} {\bibinfo {author} {\bibfnamefont {T.}~\bibnamefont
  {Morie}}, \bibinfo {author} {\bibfnamefont {T.}~\bibnamefont {Sakakibara}},
  \bibinfo {author} {\bibfnamefont {T.}~\bibnamefont {Tayama}}, \ and\ \bibinfo
  {author} {\bibfnamefont {S.}~\bibnamefont {Kunii}},\ }\href@noop {}
  {\bibfield  {journal} {\bibinfo  {journal} {J. Phys. Soc. Jpn.}\ }\textbf
  {\bibinfo {volume} {73}},\ \bibinfo {pages} {2381} (\bibinfo {year}
  {2004})}\BibitemShut {NoStop}%
\bibitem [{\citenamefont {Mannix}\ \emph {et~al.}(2005)\citenamefont {Mannix},
  \citenamefont {Tanaka}, \citenamefont {Carbone}, \citenamefont {Bernhoeft},\
  and\ \citenamefont {Kunii}}]{Mannix2005}%
  \BibitemOpen
  \bibfield  {author} {\bibinfo {author} {\bibfnamefont {D.}~\bibnamefont
  {Mannix}}, \bibinfo {author} {\bibfnamefont {Y.}~\bibnamefont {Tanaka}},
  \bibinfo {author} {\bibfnamefont {D.}~\bibnamefont {Carbone}}, \bibinfo
  {author} {\bibfnamefont {N.}~\bibnamefont {Bernhoeft}}, \ and\ \bibinfo
  {author} {\bibfnamefont {S.}~\bibnamefont {Kunii}},\ }\href@noop {}
  {\bibfield  {journal} {\bibinfo  {journal} {Phys. Rev. Lett.}\ }\textbf
  {\bibinfo {volume} {95}},\ \bibinfo {pages} {117206} (\bibinfo {year}
  {2005})}\BibitemShut {NoStop}%
\bibitem [{\citenamefont {Kuwahara}\ \emph {et~al.}(2007)\citenamefont
  {Kuwahara}, \citenamefont {Iwasa}, \citenamefont {Kohgi}, \citenamefont
  {Aso}, \citenamefont {Sera},\ and\ \citenamefont {Iga}}]{Kuwahara2007}%
  \BibitemOpen
  \bibfield  {author} {\bibinfo {author} {\bibfnamefont {K.}~\bibnamefont
  {Kuwahara}}, \bibinfo {author} {\bibfnamefont {K.}~\bibnamefont {Iwasa}},
  \bibinfo {author} {\bibfnamefont {M.}~\bibnamefont {Kohgi}}, \bibinfo
  {author} {\bibfnamefont {N.}~\bibnamefont {Aso}}, \bibinfo {author}
  {\bibfnamefont {M.}~\bibnamefont {Sera}}, \ and\ \bibinfo {author}
  {\bibfnamefont {F.}~\bibnamefont {Iga}},\ }\href@noop {} {\bibfield
  {journal} {\bibinfo  {journal} {J. Phys. Soc. Jpn.}\ }\textbf {\bibinfo
  {volume} {76}},\ \bibinfo {pages} {093702} (\bibinfo {year}
  {2007})}\BibitemShut {NoStop}%
\bibitem [{\citenamefont {Inami}\ \emph {et~al.}(2014)\citenamefont {Inami},
  \citenamefont {Michimura}, \citenamefont {Hayashi}, \citenamefont
  {Matsumura}, \citenamefont {Sera},\ and\ \citenamefont {Iga}}]{Inami2014}%
  \BibitemOpen
  \bibfield  {author} {\bibinfo {author} {\bibfnamefont {T.}~\bibnamefont
  {Inami}}, \bibinfo {author} {\bibfnamefont {S.}~\bibnamefont {Michimura}},
  \bibinfo {author} {\bibfnamefont {Y.}~\bibnamefont {Hayashi}}, \bibinfo
  {author} {\bibfnamefont {T.}~\bibnamefont {Matsumura}}, \bibinfo {author}
  {\bibfnamefont {M.}~\bibnamefont {Sera}}, \ and\ \bibinfo {author}
  {\bibfnamefont {F.}~\bibnamefont {Iga}},\ }\href@noop {} {\bibfield
  {journal} {\bibinfo  {journal} {Phys. Rev. B}\ }\textbf {\bibinfo {volume}
  {90}},\ \bibinfo {pages} {041108} (\bibinfo {year} {2014})}\BibitemShut
  {NoStop}%
\bibitem [{\citenamefont {Onimaru}\ \emph {et~al.}(2005)\citenamefont
  {Onimaru}, \citenamefont {Sakakibara}, \citenamefont {Aso}, \citenamefont
  {Yoshizawa}, \citenamefont {Suzuki},\ and\ \citenamefont
  {Takeuchi}}]{Onimaru2005}%
  \BibitemOpen
  \bibfield  {author} {\bibinfo {author} {\bibfnamefont {T.}~\bibnamefont
  {Onimaru}}, \bibinfo {author} {\bibfnamefont {T.}~\bibnamefont {Sakakibara}},
  \bibinfo {author} {\bibfnamefont {N.}~\bibnamefont {Aso}}, \bibinfo {author}
  {\bibfnamefont {H.}~\bibnamefont {Yoshizawa}}, \bibinfo {author}
  {\bibfnamefont {H.~S.}\ \bibnamefont {Suzuki}}, \ and\ \bibinfo {author}
  {\bibfnamefont {T.}~\bibnamefont {Takeuchi}},\ }\href@noop {} {\bibfield
  {journal} {\bibinfo  {journal} {Phys. Rev. Lett.}\ }\textbf {\bibinfo
  {volume} {94}},\ \bibinfo {pages} {197201} (\bibinfo {year}
  {2005})}\BibitemShut {NoStop}%
\bibitem [{\citenamefont {Onimaru}\ \emph {et~al.}(2010)\citenamefont
  {Onimaru}, \citenamefont {Matsumoto}, \citenamefont {Inoue}, \citenamefont
  {Umeo}, \citenamefont {Saiga}, \citenamefont {Matsushita}, \citenamefont
  {Tamura}, \citenamefont {Nishimoto}, \citenamefont {Ishii}, \citenamefont
  {Suzuki},\ and\ \citenamefont {Takabatake}}]{Onimaru2010}%
  \BibitemOpen
  \bibfield  {author} {\bibinfo {author} {\bibfnamefont {T.}~\bibnamefont
  {Onimaru}}, \bibinfo {author} {\bibfnamefont {K.~T.}\ \bibnamefont
  {Matsumoto}}, \bibinfo {author} {\bibfnamefont {Y.~F.}\ \bibnamefont
  {Inoue}}, \bibinfo {author} {\bibfnamefont {K.}~\bibnamefont {Umeo}},
  \bibinfo {author} {\bibfnamefont {Y.}~\bibnamefont {Saiga}}, \bibinfo
  {author} {\bibfnamefont {Y.}~\bibnamefont {Matsushita}}, \bibinfo {author}
  {\bibfnamefont {R.}~\bibnamefont {Tamura}}, \bibinfo {author} {\bibfnamefont
  {K.}~\bibnamefont {Nishimoto}}, \bibinfo {author} {\bibfnamefont
  {I.}~\bibnamefont {Ishii}}, \bibinfo {author} {\bibfnamefont
  {T.}~\bibnamefont {Suzuki}}, \ and\ \bibinfo {author} {\bibfnamefont
  {T.}~\bibnamefont {Takabatake}},\ }\href@noop {} {\bibfield  {journal}
  {\bibinfo  {journal} {J. Phys. Soc. Jpn.}\ }\textbf {\bibinfo {volume}
  {79}},\ \bibinfo {pages} {033704} (\bibinfo {year} {2010})}\BibitemShut
  {NoStop}%
\bibitem [{\citenamefont {Sakai}\ \emph {et~al.}(2012)\citenamefont {Sakai},
  \citenamefont {Kuga},\ and\ \citenamefont {Nakatsuji}}]{Sakai2012}%
  \BibitemOpen
  \bibfield  {author} {\bibinfo {author} {\bibfnamefont {A.}~\bibnamefont
  {Sakai}}, \bibinfo {author} {\bibfnamefont {K.}~\bibnamefont {Kuga}}, \ and\
  \bibinfo {author} {\bibfnamefont {S.}~\bibnamefont {Nakatsuji}},\ }\href@noop
  {} {\bibfield  {journal} {\bibinfo  {journal} {J. Phys. Soc. Jpn.}\ }\textbf
  {\bibinfo {volume} {81}},\ \bibinfo {pages} {083702} (\bibinfo {year}
  {2012})}\BibitemShut {NoStop}%
\bibitem [{\citenamefont {Matsubayashi}\ \emph {et~al.}(2012)\citenamefont
  {Matsubayashi}, \citenamefont {Tanaka}, \citenamefont {Sakai}, \citenamefont
  {Nakatsuji}, \citenamefont {Kubo},\ and\ \citenamefont
  {Uwatoko}}]{Matsubayashi2012}%
  \BibitemOpen
  \bibfield  {author} {\bibinfo {author} {\bibfnamefont {K.}~\bibnamefont
  {Matsubayashi}}, \bibinfo {author} {\bibfnamefont {T.}~\bibnamefont
  {Tanaka}}, \bibinfo {author} {\bibfnamefont {A.}~\bibnamefont {Sakai}},
  \bibinfo {author} {\bibfnamefont {S.}~\bibnamefont {Nakatsuji}}, \bibinfo
  {author} {\bibfnamefont {Y.}~\bibnamefont {Kubo}}, \ and\ \bibinfo {author}
  {\bibfnamefont {Y.}~\bibnamefont {Uwatoko}},\ }\href@noop {} {\bibfield
  {journal} {\bibinfo  {journal} {Phys. Rev. Lett.}\ }\textbf {\bibinfo
  {volume} {109}},\ \bibinfo {pages} {187004} (\bibinfo {year}
  {2012})}\BibitemShut {NoStop}%
\bibitem [{\citenamefont {Onimaru}\ \emph {et~al.}(2012)\citenamefont
  {Onimaru}, \citenamefont {Nagasawa}, \citenamefont {Matsumoto}, \citenamefont
  {Wakiya}, \citenamefont {Umeo}, \citenamefont {Kittaka}, \citenamefont
  {Sakakibara}, \citenamefont {Matsushita},\ and\ \citenamefont
  {Takabatake}}]{Onimaru2012}%
  \BibitemOpen
  \bibfield  {author} {\bibinfo {author} {\bibfnamefont {T.}~\bibnamefont
  {Onimaru}}, \bibinfo {author} {\bibfnamefont {N.}~\bibnamefont {Nagasawa}},
  \bibinfo {author} {\bibfnamefont {K.~T.}\ \bibnamefont {Matsumoto}}, \bibinfo
  {author} {\bibfnamefont {K.}~\bibnamefont {Wakiya}}, \bibinfo {author}
  {\bibfnamefont {K.}~\bibnamefont {Umeo}}, \bibinfo {author} {\bibfnamefont
  {S.}~\bibnamefont {Kittaka}}, \bibinfo {author} {\bibfnamefont
  {T.}~\bibnamefont {Sakakibara}}, \bibinfo {author} {\bibfnamefont
  {Y.}~\bibnamefont {Matsushita}}, \ and\ \bibinfo {author} {\bibfnamefont
  {T.}~\bibnamefont {Takabatake}},\ }\href@noop {} {\bibfield  {journal}
  {\bibinfo  {journal} {Phys. Rev. B}\ }\textbf {\bibinfo {volume} {86}},\
  \bibinfo {pages} {184426} (\bibinfo {year} {2012})}\BibitemShut {NoStop}%
\bibitem [{\citenamefont {Tsujimoto}\ \emph {et~al.}(2014)\citenamefont
  {Tsujimoto}, \citenamefont {Matsumoto}, \citenamefont {Tomita}, \citenamefont
  {Sakai},\ and\ \citenamefont {Nakatsuji}}]{Tsujimoto2014}%
  \BibitemOpen
  \bibfield  {author} {\bibinfo {author} {\bibfnamefont {M.}~\bibnamefont
  {Tsujimoto}}, \bibinfo {author} {\bibfnamefont {Y.}~\bibnamefont
  {Matsumoto}}, \bibinfo {author} {\bibfnamefont {T.}~\bibnamefont {Tomita}},
  \bibinfo {author} {\bibfnamefont {A.}~\bibnamefont {Sakai}}, \ and\ \bibinfo
  {author} {\bibfnamefont {S.}~\bibnamefont {Nakatsuji}},\ }\href@noop {}
  {\bibfield  {journal} {\bibinfo  {journal} {Phys. Rev. Lett.}\ }\textbf
  {\bibinfo {volume} {113}},\ \bibinfo {pages} {267001} (\bibinfo {year}
  {2014})}\BibitemShut {NoStop}%
\bibitem [{\citenamefont {Onimaru}\ and\ \citenamefont
  {Kusunose}(2016)}]{Onimaru2016}%
  \BibitemOpen
  \bibfield  {author} {\bibinfo {author} {\bibfnamefont {T.}~\bibnamefont
  {Onimaru}}\ and\ \bibinfo {author} {\bibfnamefont {H.}~\bibnamefont
  {Kusunose}},\ }\href@noop {} {\bibfield  {journal} {\bibinfo  {journal} {J.
  Phys. Soc. Jpn.}\ }\textbf {\bibinfo {volume} {85}},\ \bibinfo {pages}
  {082002} (\bibinfo {year} {2016})}\BibitemShut {NoStop}%
\bibitem [{\citenamefont {Hotta}\ and\ \citenamefont
  {Harima}(2006)}]{Hotta2006}%
  \BibitemOpen
  \bibfield  {author} {\bibinfo {author} {\bibfnamefont {T.}~\bibnamefont
  {Hotta}}\ and\ \bibinfo {author} {\bibfnamefont {H.}~\bibnamefont {Harima}},\
  }\href@noop {} {\bibfield  {journal} {\bibinfo  {journal} {J. Phys. Soc.
  Jpn.}\ }\textbf {\bibinfo {volume} {75}},\ \bibinfo {pages} {124711}
  (\bibinfo {year} {2006})}\BibitemShut {NoStop}%
\bibitem [{\citenamefont {Hotta}\ and\ \citenamefont {Ueda}(2003)}]{Hotta2003}%
  \BibitemOpen
  \bibfield  {author} {\bibinfo {author} {\bibfnamefont {T.}~\bibnamefont
  {Hotta}}\ and\ \bibinfo {author} {\bibfnamefont {K.}~\bibnamefont {Ueda}},\
  }\href@noop {} {\bibfield  {journal} {\bibinfo  {journal} {Phys. Rev. B}\
  }\textbf {\bibinfo {volume} {67}},\ \bibinfo {pages} {104518} (\bibinfo
  {year} {2003})}\BibitemShut {NoStop}%
\bibitem [{\citenamefont {Shiina}\ \emph {et~al.}(1997)\citenamefont {Shiina},
  \citenamefont {Nishitani},\ and\ \citenamefont {Shiba}}]{Shiina1997}%
  \BibitemOpen
  \bibfield  {author} {\bibinfo {author} {\bibfnamefont {R.}~\bibnamefont
  {Shiina}}, \bibinfo {author} {\bibfnamefont {T.}~\bibnamefont {Nishitani}}, \
  and\ \bibinfo {author} {\bibfnamefont {H.}~\bibnamefont {Shiba}},\
  }\href@noop {} {\bibfield  {journal} {\bibinfo  {journal} {J. Phys. Soc.
  Jpn.}\ }\textbf {\bibinfo {volume} {66}},\ \bibinfo {pages} {3159} (\bibinfo
  {year} {1997})}\BibitemShut {NoStop}%
\bibitem [{\citenamefont {van~den Brink}\ \emph {et~al.}(1999)\citenamefont
  {van~den Brink}, \citenamefont {Horsch}, \citenamefont {Mack},\ and\
  \citenamefont {Ole{\'{s}}}}]{Brink1999}%
  \BibitemOpen
  \bibfield  {author} {\bibinfo {author} {\bibfnamefont {J.}~\bibnamefont
  {van~den Brink}}, \bibinfo {author} {\bibfnamefont {P.}~\bibnamefont
  {Horsch}}, \bibinfo {author} {\bibfnamefont {F.}~\bibnamefont {Mack}}, \ and\
  \bibinfo {author} {\bibfnamefont {A.~M.}\ \bibnamefont {Ole{\'{s}}}},\
  }\href@noop {} {\bibfield  {journal} {\bibinfo  {journal} {Phys. Rev. B}\
  }\textbf {\bibinfo {volume} {59}},\ \bibinfo {pages} {6795} (\bibinfo {year}
  {1999})}\BibitemShut {NoStop}%
\bibitem [{\citenamefont {Ishihara}\ and\ \citenamefont
  {Maekawa}(2000)}]{Ishihara2000}%
  \BibitemOpen
  \bibfield  {author} {\bibinfo {author} {\bibfnamefont {S.}~\bibnamefont
  {Ishihara}}\ and\ \bibinfo {author} {\bibfnamefont {S.}~\bibnamefont
  {Maekawa}},\ }\href@noop {} {\bibfield  {journal} {\bibinfo  {journal} {Phys.
  Rev. B}\ }\textbf {\bibinfo {volume} {62}},\ \bibinfo {pages} {2338}
  (\bibinfo {year} {2000})}\BibitemShut {NoStop}%
\bibitem [{\citenamefont {Kubo}(2002)}]{Kubo2002JPSJNo5}%
  \BibitemOpen
  \bibfield  {author} {\bibinfo {author} {\bibfnamefont {K.}~\bibnamefont
  {Kubo}},\ }\href@noop {} {\bibfield  {journal} {\bibinfo  {journal} {J. Phys.
  Soc. Jpn.}\ }\textbf {\bibinfo {volume} {71}},\ \bibinfo {pages} {1308}
  (\bibinfo {year} {2002})}\BibitemShut {NoStop}%
\bibitem [{\citenamefont {Onimaru}\ \emph {et~al.}(2004)\citenamefont
  {Onimaru}, \citenamefont {Sakakibara}, \citenamefont {Harita}, \citenamefont
  {Tayama}, \citenamefont {Aoki},\ and\ \citenamefont
  {{\=O}nuki}}]{Onimaru2004}%
  \BibitemOpen
  \bibfield  {author} {\bibinfo {author} {\bibfnamefont {T.}~\bibnamefont
  {Onimaru}}, \bibinfo {author} {\bibfnamefont {T.}~\bibnamefont {Sakakibara}},
  \bibinfo {author} {\bibfnamefont {A.}~\bibnamefont {Harita}}, \bibinfo
  {author} {\bibfnamefont {T.}~\bibnamefont {Tayama}}, \bibinfo {author}
  {\bibfnamefont {D.}~\bibnamefont {Aoki}}, \ and\ \bibinfo {author}
  {\bibfnamefont {Y.}~\bibnamefont {{\=O}nuki}},\ }\href@noop {} {\bibfield
  {journal} {\bibinfo  {journal} {J. Phys. Soc. Jpn.}\ }\textbf {\bibinfo
  {volume} {73}},\ \bibinfo {pages} {2377} (\bibinfo {year}
  {2004})}\BibitemShut {NoStop}%
\bibitem [{\citenamefont {Takahashi}\ and\ \citenamefont
  {Shiba}(2000)}]{Takahashi2000}%
  \BibitemOpen
  \bibfield  {author} {\bibinfo {author} {\bibfnamefont {A.}~\bibnamefont
  {Takahashi}}\ and\ \bibinfo {author} {\bibfnamefont {H.}~\bibnamefont
  {Shiba}},\ }\href@noop {} {\bibfield  {journal} {\bibinfo  {journal} {J.
  Phys. Soc. Jpn.}\ }\textbf {\bibinfo {volume} {69}},\ \bibinfo {pages} {3328}
  (\bibinfo {year} {2000})}\BibitemShut {NoStop}%
\bibitem [{\citenamefont {Maezono}\ and\ \citenamefont
  {Nagaosa}(2000)}]{Maezono2000}%
  \BibitemOpen
  \bibfield  {author} {\bibinfo {author} {\bibfnamefont {R.}~\bibnamefont
  {Maezono}}\ and\ \bibinfo {author} {\bibfnamefont {N.}~\bibnamefont
  {Nagaosa}},\ }\href@noop {} {\bibfield  {journal} {\bibinfo  {journal} {Phys.
  Rev. B}\ }\textbf {\bibinfo {volume} {62}},\ \bibinfo {pages} {11576}
  (\bibinfo {year} {2000})}\BibitemShut {NoStop}%
\bibitem [{\citenamefont {van~den Brink}\ and\ \citenamefont
  {Khomskii}(2001)}]{Brink2001}%
  \BibitemOpen
  \bibfield  {author} {\bibinfo {author} {\bibfnamefont {J.}~\bibnamefont
  {van~den Brink}}\ and\ \bibinfo {author} {\bibfnamefont {D.}~\bibnamefont
  {Khomskii}},\ }\href@noop {} {\bibfield  {journal} {\bibinfo  {journal}
  {Phys. Rev. B}\ }\textbf {\bibinfo {volume} {63}},\ \bibinfo {pages}
  {140416(R)} (\bibinfo {year} {2001})}\BibitemShut {NoStop}%
\bibitem [{\citenamefont {Khomskii}(2001)}]{Khomskii2001}%
  \BibitemOpen
  \bibfield  {author} {\bibinfo {author} {\bibfnamefont {D.~I.}\ \bibnamefont
  {Khomskii}},\ }\href@noop {} {\bibfield  {journal} {\bibinfo  {journal} {Int.
  J. Mod. Phys. B}\ }\textbf {\bibinfo {volume} {15}},\ \bibinfo {pages} {2665}
  (\bibinfo {year} {2001})}\BibitemShut {NoStop}%
\bibitem [{\citenamefont {Kubo}\ and\ \citenamefont
  {Hirashima}(2002)}]{Kubo2002JPSJNo1}%
  \BibitemOpen
  \bibfield  {author} {\bibinfo {author} {\bibfnamefont {K.}~\bibnamefont
  {Kubo}}\ and\ \bibinfo {author} {\bibfnamefont {D.~S.}\ \bibnamefont
  {Hirashima}},\ }\href@noop {} {\bibfield  {journal} {\bibinfo  {journal} {J.
  Phys. Soc. Jpn.}\ }\textbf {\bibinfo {volume} {71}},\ \bibinfo {pages} {183}
  (\bibinfo {year} {2002})}\BibitemShut {NoStop}%
\bibitem [{\citenamefont {Sakai}\ \emph {et~al.}(2003)\citenamefont {Sakai},
  \citenamefont {Shiina},\ and\ \citenamefont {Shiba}}]{Sakai2003}%
  \BibitemOpen
  \bibfield  {author} {\bibinfo {author} {\bibfnamefont {O.}~\bibnamefont
  {Sakai}}, \bibinfo {author} {\bibfnamefont {R.}~\bibnamefont {Shiina}}, \
  and\ \bibinfo {author} {\bibfnamefont {H.}~\bibnamefont {Shiba}},\
  }\href@noop {} {\bibfield  {journal} {\bibinfo  {journal} {J. Phys. Soc.
  Jpn.}\ }\textbf {\bibinfo {volume} {72}},\ \bibinfo {pages} {1534} (\bibinfo
  {year} {2003})}\BibitemShut {NoStop}%
\end{thebibliography}

%

\end{document}